\documentclass[preprint,amsmath,amssymb,aps,showkeys,showpacs]{revtex4}
\usepackage[english]{babel}
\usepackage{graphicx}
\usepackage{graphics}
\usepackage{amsmath}
\usepackage{dcolumn}
\usepackage{amssymb}
\usepackage{bm}

\begin{document}
\title{Effects of rotation in the spacetime with a distortion of a vertical line into a vertical spiral}
\author{K. Bakke}
\email{kbakke@fisica.ufpb.br}
\affiliation{Departamento de F\'isica, Universidade Federal da Para\'iba, Caixa Postal 5008, 58051-900, Jo\~ao Pessoa, PB, Brazil.}

\begin{abstract}

It is investigated the effects of rotation on the scalar field in the spacetime with the distortion of a vertical line into a vertical spiral. By analysing the upper limit of the radial coordinate that stems from the effects of rotation and the topology of the defect, it is considered this upper limit of the radial coordinate as a boundary condition analogous to a hard-wall confining potential. Then, it is obtained a relativistic spectrum of energy in a particular case. In addition, it is analysed a relativistic analogue of the Aharonov-Bohm effect for bound states.

\end{abstract}

\keywords{Aharonov-Bohm effect, linear topological defects, screw dislocation, rotating reference frame, radial Mathieu equation}

\maketitle

\section{Introduction}

The possibility of arising topological defects from phase transitions during the universe evolution has raised discussions about spacetimes characterized by the presence of torsion and curvature. The best known example is the cosmic string spacetime \cite{av,vil,JSD,kibble,staro,put,fur10}. In connection with the description of topological defects in solids \cite{kat,kleinert}, it has been considered spacetimes with a screw dislocation \cite{put,valdir,vb} and a spiral dislocation \cite{bf,mb2,vb}. In recent years, the effects associated with these topological defect spacetimes have been investigated on the interface between general relativity and quantum mechanics \cite{valdir,vb,mb2,b3,valdir2,valdir3,bf,1,2,3,4,5,6,7,8,9,10,11,12,13,14,15,16,17,18,19,20,21,22,23,24,25,26,27,28,29,30,31}. Recently, G\"odel-type spacetimes with topological defects have been analysed in this interface \cite{g1,g2,g3,g4,g5,g6,g7,g8,g9}.

Another perspective was given by considering a uniformly rotating reference frame. From the classical mechanics point of view, it has been raised that topological defects can determine at least the upper limit of the radial coordinate in this rotating frame \cite{vb,mb2,b3}. This characteristic of the topological defect spacetimes in a uniformly rotating reference frame has agreed with the point raised by Landau and Lifshitz \cite{landau3}, where they showed that the Minkowski spacetime has a singular behaviour at larges distances in a uniformly rotating frame. On the other hand, by going through quantum systems in backgrounds with topological defects and rotation, quantum effects have been reported in the literature through geometric quantum phases \cite{bf3}. Besides, the upper limit of the radial coordinate that stems from the uniformly rotating reference frame has played the role of a boundary condition on the wave function. This has drawn attention to the influence of rotation on the relativistic spectrum of energy from a geometric point of view \cite{b2,vb,mb2}.

In this work, we extend the description of a topological defect in a solid to the context of gravitation. We introduce a spacetime with the distortion of a vertical line into a vertical spiral. Then, we consider a uniformly rotating reference frame. Thereby, we start by analysing the effects of rotation and the topology of the defect on the upper limit of the radial coordinate. Further, we consider this upper limit of the radial coordinate as a boundary condition analogous to a hard-wall confining potential, and thus, we obtain the relativistic spectrum of energy in a particular case. Besides, we analyse the possibility of finding an analogue of the Aharonov-Bohm effect for bound states \cite{pesk}.

This paper is structured as follows: in section II, we introduce the spacetime with the distortion of a vertical line into a vertical spiral. We thus consider a uniformly rotating reference frame and obtain the upper limit of the radial coordinate. Then, by using this upper limit of the radial coordinate as a boundary condition analogous to a hard-wall confining potential, we search for relativistic bound states solutions to the Klein-Gordon equation in this topological defect spacetime. Finally, we analyse the Aharonov-Bohm-type effect for bound states; in section III, we present our conclusions.

\section{Effects of rotation on the scalar field in in the spacetime with a distortion of a vertical line into a vertical spiral}

In Ref. \cite{val} is shown that an elastic medium with the distortion of a vertical line into a vertical spiral can be described by a metric tensor in agreement with the Katanaev-Volovich approach \cite{kat}. In this work, we extend this discussion to the context of gravitation. In this way, we consider a spacetime with the distortion of a vertical line into a vertical spiral by describing it with the line element (with $\hbar=1$ and $c=1$):
\begin{eqnarray}
ds^{2}=-dt^{2}+dr^{2}+r^{2}d\varphi^{2}+2\beta\,d\varphi\,dz+dz^{2},
\label{1.1}
\end{eqnarray}
where $0\,<\,r\,<\,\infty$, $0\leq\varphi\leq2\pi$ and $-\infty\,<\,z\,<\,\infty$. The parameter $\beta$ is a constant that characterizes the torsion field (dislocation) in the spacetime. It can be defined in the range $0\,<\,\beta\,<\,1$. Next, let us consider uniformly rotating frame. Thereby, let us perform the coordinate transformation: $\varphi\rightarrow\varphi+\omega\,t$. Then, the line element (\ref{1.1}) becomes 
\begin{eqnarray}
ds^{2}=-\left(1-\omega^{2}r^{2}\right)dt^{2}+2\omega\,r^{2}\,d\varphi\,dt+2\omega\beta\,dt\,dz+dr^{2}+r^{2}d\varphi^{2}+2\beta\,d\varphi\,dz+dz^{2}.
\label{1.2}
\end{eqnarray}

Hence, from Eq. (\ref{1.2}), we have that the line element has a singular behaviour at larges distances because the radial coordinate is defined in the range:
\begin{eqnarray}
0\leq\,r\,<1/\omega.
\label{1.2a}
\end{eqnarray}
Despite having the presence of torsion in the spacetime with the distortion of a vertical line into a vertical spiral, there is no influence of torsion on the range (\ref{1.2a}). By contrast, it is shown in Ref. \cite{vb} that there is the influence of torsion on the range of the possible values of the radial coordinate in the spacetime with the distortion of a circular curve into a vertical spiral and also in the spacetime with a spiral dislocation.

Our aim is to investigate rotating effects on the scalar field subject to a hard-wall confining potential in the spacetime with the distortion of a vertical line into a vertical spiral. According to Refs. \cite{valdir,valdir3,vb}, the Klein-Gordon equation in the spacetime with a topological defect can be written in the form: 
\begin{eqnarray}
m^{2}\Phi=\frac{1}{\sqrt{-g}}\,\partial_{\mu}\left(g^{\mu\nu}\,\sqrt{-g}\,\partial_{\nu}\right)\Phi,
\label{1.3}
\end{eqnarray}
where $g_{\mu\nu}$ is the metric tensor, $g^{\mu\nu}$ is the inverse of $g_{\mu\nu}$ and $g=\mathrm{det}\left|g_{\mu\nu}\right|$. Observe that the indices $\left\{i,\,j\right\}$ run over the space coordinates. Thereby, with the line element (\ref{1.2}), the Klein-Gordon equation has the form:
\begin{eqnarray}
m^{2}\Phi=-\left[\frac{\partial}{\partial t}-\omega\frac{\partial}{\partial\varphi}\right]^{2}\Phi+\frac{\partial^{2}\Phi}{\partial r^{2}}+\frac{r}{\left(r^{2}-\beta^{2}\right)}\,\frac{\partial\Phi}{\partial r}+\frac{1}{\left(r^{2}-\beta^{2}\right)}\left[\frac{\partial}{\partial\varphi}-\beta\frac{\partial}{\partial z}\right]^{2}\Phi+\frac{\partial^{2}\Phi}{\partial z^{2}}.
\label{1.4}
\end{eqnarray}

Observe that the quantum operators $\hat{p}_{z}=-i\partial_{z}$ and $\hat{L}_{z}=-i\partial_{\varphi}$ commutes with the Hamiltonian operator given in the right-hand side of Eq. (\ref{1.4}). Therefore, a possible way of writing the solution to Eq. (\ref{1.4}) is in terms of the eigenvalues of these operators. In this way, let us write $\Phi\left(t,\,r,\,\varphi,\,z\right)=e^{-i\mathcal{E}t+il\varphi+ikz}\,f\left(r\right)$, where $k=\mathrm{const}$ and $l=0,\pm1,\pm2,\pm3\ldots$ are the eigenvalues of the operators $\hat{p}_{z}=-i\partial_{z}$ and $\hat{L}_{z}=-i\partial_{\varphi}$, respectively. With this solution, we obtain the following radial equation:
\begin{eqnarray}
f''+\frac{r}{\left(r^{2}-\beta^{2}\right)}\,f'-\frac{\gamma^{2}}{\left(r^{2}-\beta^{2}\right)}\,f+\left[\left(\mathcal{E}+\omega\,l\right)^{2}-m^{2}-k^{2}\right]\,f=0,
\label{1.5}
\end{eqnarray}
where $\gamma=\left(l-\beta\,k\right)$. Let us proceed our discussion by defining a dimensionless parameter $y$ as
\begin{eqnarray}
\cosh y=\frac{r}{\beta},
\label{1.6}
\end{eqnarray}
therefore, the radial equation (\ref{1.5}) becomes
\begin{eqnarray}
f''+\left[2\,q^{2}\,\cosh\left(2y\right)-\lambda\right]f=0,
\label{1.7}
\end{eqnarray}
where we have defined the following parameters in Eq. (\ref{1.7}):
\begin{eqnarray}
\lambda&=&\gamma^{2}+\frac{\beta^{2}}{2}\left[\left(\mathcal{E}+\omega\,l\right)^{2}-m^{2}-k^{2}\right];\nonumber\\
[-2mm]\label{1.8}\\[-2mm]
q^{2}&=&\frac{\beta^{2}}{2}\left[\left(\mathcal{E}+\omega\,l\right)^{2}-m^{2}-k^{2}\right].\nonumber
\end{eqnarray}
It is worth observing that Eq. (\ref{1.7}) is called in the literature as the modified Mathieu equation or radial Mathieu equation \cite{abra,arf,mat1,mat2}. With the purpose of simplifying our analysis, let us follow Refs. \cite{mat2,sb} and define the parameter:
\begin{eqnarray}
x&=&2q\,\cosh y\nonumber\\
&=&\beta\sqrt{\left[\left(\mathcal{E}+\omega\,l\right)^{2}-m^{2}-k^{2}\right]}\,\cosh y.
\label{1.8}
\end{eqnarray}
Then, by substituting Eq. (\ref{1.8}) into Eq. (\ref{1.7}), the modified Mathieu equation becomes:
\begin{eqnarray}
f''+\frac{1}{x}\,f'+f-\frac{1}{x^{2}}\left[\lambda\,f+2q^{2}\left(f+2f''\right)\right]=0.
\label{1.9}
\end{eqnarray}

Since the parameter $\beta$ that characterizes the topological defect in the line element (\ref{1.1}) or (\ref{1.2}) is defined in $0<\,\beta\,<\,1$, we can neglect the terms proportional to $\beta^{2}$, without loss of generality. Thereby, we can write the last term of Eq. (\ref{1.9}) as
\begin{eqnarray}
\lambda\,f+2q^{2}\left(f+2f''\right)\approx\gamma^{2}\,f.
\label{1.10}
\end{eqnarray}
By using the approximation given in Eq. (\ref{1.10}), hence, Eq. (\ref{1.9}) becomes
\begin{eqnarray}
f''+\frac{1}{x}\,f'-\frac{\gamma^{2}}{x^{2}}\,f+f=0,
\label{1.11}
\end{eqnarray}
which is the Bessel differential equation \cite{arf,abra}. Observe in the line element (\ref{1.2}) that the spacetime has the cylindrical symmetry. Thereby, we search for a regular solution at the origin. In this perspective, a regular solution to Eq. (\ref{1.11}), when $r=0\Rightarrow x=0$, is given by
\begin{eqnarray}
f\left(x\right)=A\,J_{\left|\gamma\right|}\left(x\right),
\label{1.12}
\end{eqnarray} 
where $J_{\left|\gamma\right|}\left(x\right)$ is the Bessel function of first kind \cite{abra,arf} and $A$ is a constant.

Returning to the line element (\ref{1.2}), we have that the radial coordinate is restricted to the range determined in Eq. (\ref{1.2a}). Thereby, the radial wave function of the scalar particle must vanish when $r\rightarrow r_{0}=1/\omega$, i.e., when $x\rightarrow x_{0}$, where $x_{0}$ is given by using the relations $r_{0}=1/\omega$ and Eqs. (\ref{1.6}) and (\ref{1.8}):
\begin{eqnarray}
x_{0}&=&\beta\sqrt{\left[\left(\mathcal{E}+\omega\,l\right)^{2}-m^{2}-k^{2}\right]}\,\cosh y_{0}\nonumber\\
&=&\frac{1}{\omega}\,\sqrt{\left[\left(\mathcal{E}+\omega\,l\right)^{2}-m^{2}-k^{2}\right]}.
\label{1.13}
\end{eqnarray}
This means that the radial wave function of the scalar particle must satisfy the boundary condition:
\begin{eqnarray}
f\left(x_{0}\right)=0.
\label{1.14}
\end{eqnarray}
Hence, the boundary condition (\ref{1.14}) corresponds to confinement of the scalar field to a hard-wall confining potential. Besides, since $x_{0}$ is determined by the restriction on the radial coordinate given in Eq. (\ref{1.2a}), thus, we have that the geometry of the spacetime with the distortion of a vertical line into a vertical spiral plays the role of this hard-wall confining potential in the uniformly rotating frame even though no influence of torsion exists on the upper limit given in (\ref{1.2a}). This behaviour of having the geometry of the spacetime playing the role of a hard-wall confining potential in uniformly rotating frame has been analysed in the Minkowski spacetime \cite{b2} and other topological defect spacetimes \cite{mb2,vb,b3}.

Let us go further by considering a particular case of the Bessel function. Let us assume that $x_{0}\gg1$. Then, when $\left|\gamma\right|$ is fixed and  $x_{0}\gg1$, the Bessel function can be written as \cite{abra,valdir}:
\begin{eqnarray}
J_{\left|\gamma\right|}\left(x_{0}\right)\rightarrow\sqrt{\frac{2}{\pi\,x_{0}}}\,\cos\left(x_{0}-\frac{\left|\gamma\right|\,\pi}{2}-\frac{\pi}{4}\right).
\label{1.15}
\end{eqnarray}

Therefore, by substituting (\ref{1.15}) into (\ref{1.12}), we obtain with the boundary condition (\ref{1.14}):
\begin{eqnarray}
\mathcal{E}_{n,\,l,\,k}\approx-\omega\,l\pm\sqrt{m^{2}+k^{2}+\pi^{2}\omega^{2}\left[n+\frac{1}{2}\left|l-\beta\,k\right|+\frac{3}{4}\right]^{2}},
\label{1.16}
\end{eqnarray}
where $n=0,1,2,\ldots$ is the quantum number related to the radial modes and $l=0,\pm1,\pm2,\ldots$ is the angular momentum quantum number. 

The discrete spectrum of energy given in Eq. (\ref{1.16}) is obtained from the confinement of the scalar field to a hard-wall confining potential. We can observe that there is the influence of rotation and the topology of the spacetime with the distortion of a vertical line into a vertical spiral on the relativistic energy levels (\ref{1.16}). The contribution that stems from the topology of the spacetime is given by the the effective angular momentum $\gamma=\left(l-\beta\,k\right)$. This shift in the angular momentum quantum number occurs even though no interaction between the quantum particle and the topological defect exists. Therefore, it corresponds to an analogue of the Aharonov-Bohm effect for bound states \cite{pesk,valdir,valdir3,vb,fur3}. On the other hand, one of the contributions that stems from the effects of rotation is given by the presence of the fixed radius $r_{0}=1/\omega$ in the second term of the right-hand side of Eq. (\ref{1.16}). The second contribution that stems from the effects of rotation is given by the first term of the right-hand side of Eq. (\ref{1.16}), i.e., the coupling between the angular velocity $\omega$ and the angular momentum quantum number $l$. It gives rise to a Sagnac-type effect \cite{sag,sag2,sag5,r1,r2,r3,vb}. 

Furthermore, by taking $\beta=0$, the relativistic energy levels (\ref{1.16}) becomes
\begin{eqnarray}
\mathcal{E}_{n,\,l,\,k}\approx-\omega\,l\pm\sqrt{m^{2}+k^{2}+\pi^{2}\omega^{2}\left[n+\frac{\left|l\right|}{2}+\frac{3}{4}\right]^{2}}.
\label{1.17}
\end{eqnarray}
Therefore, we recover the relativistic energy levels for a scalar field subject to a hard-wall confining potential in the Minkowski spacetime in a uniformly rotating frame \cite{vb}.


\section{Conclusions}

We have analysed effects of rotation on the scalar field in the spacetime with a distortion of a vertical line into a vertical spiral. We have seen, despite the presence of torsion in the spacetime, there is no contribution of the topology of the spacetime on the possible values of the radial coordinate in the uniformly rotating frame. The restriction on the radial coordinate depends only on the angular velocity of the rotating frame as shown in Eq. (\ref{1.2a}).

Besides, in the analysis of the radial equation, we have focused on the case where the terms proportional to $\beta^{2}$ could be neglected. We have seen that the solution to the radial equation can be written in terms of the Bessel function. Then, we have considered a boundary condition where the wave function vanishes when $r\rightarrow r_{0}=1/\omega$. This is analogous to the confinement of the scalar field to a hard-wall confining potential. In this case, the geometry of the spacetime with the distortion of a vertical line into a vertical spiral plays the role of the hard-wall confining potential in the uniformly rotating frame even though no influence of torsion exists on the upper limit given in (\ref{1.2a}). Then, by analysing the asymptotic expression of the Bessel function, we have obtained a discrete spectrum of energy. We have observed in these relativistic energy levels that there exists the influence of rotation and the topology of the spacetime with the distortion of a vertical line into a vertical spiral. The contribution of the topology of the spacetime yields a shift in the angular momentum quantum number that corresponds to an analogue of the Aharonov-Bohm effect for bound states. Furthermore, the effects of rotation yield the presence of the fixed radius $r_{0}=1/\omega$ and the coupling between the angular velocity $\omega$ and the angular momentum quantum number $l$ (a Sagnac-type effect).

Finally, we have shown that, by taking $\beta=0$, we can recover the relativistic energy levels for a scalar field subject to a hard-wall confining potential in the Minkowski spacetime in a uniformly rotating frame \cite{vb}.


\acknowledgments{The author would like to thank the Brazilian agency CNPq for financial support.}

\end{document}